\documentclass[a4paper, 10pt, conference]{IEEEtran}

\usepackage{url}
\usepackage{graphicx}
\usepackage{amsfonts}
\usepackage{flushend} 
\usepackage{caption}
\usepackage{subcaption}
\usepackage{color}
\usepackage{stfloats}
\usepackage{color,soul}
\usepackage{amsmath}
\usepackage{multirow}
\usepackage{hhline}
\usepackage{gensymb}

\IEEEoverridecommandlockouts                              

 \title{\LARGE \bf
 Enabling End-to-End Secure Connectivity for Low-Power IoT Devices with UAVs
 }

\author{\IEEEauthorblockN{Archana Rajakaruna\IEEEauthorrefmark{1}, Ahsan Manzoor\IEEEauthorrefmark{1},
Pawani Porambage\IEEEauthorrefmark{1},
Madhusanka Liyanage\IEEEauthorrefmark{1}\IEEEauthorrefmark{3}, \\ 
Mika Ylianttila\IEEEauthorrefmark{1}, Andrei Gurtov\IEEEauthorrefmark{2}}
\IEEEauthorblockA{\IEEEauthorrefmark{1}Centre for Wireless Communications, University of Oulu, Finland \\
\IEEEauthorrefmark{3}School of Computer Science, University College Dublin, Ireland. \\
\IEEEauthorrefmark{2}Department of Computer and Information Science, Link\"oping University, Sweden}
\IEEEauthorblockA{\IEEEauthorrefmark{1}\{firstname.lastname\}@oulu.fi, \IEEEauthorrefmark{3}madhusanka@ucd.ie \IEEEauthorrefmark{2}gurtov@acm.org}
}

\begin{document}

\maketitle
\thispagestyle{empty}
\pagestyle{empty}

\begin{abstract}
The proliferation of the Internet of Things (IoT) technologies have strengthen the self-monitoring and autonomous characteristics of the sensor networks deployed in numerous application areas. The recent developments of the edge computing paradigms have also enabled on-site processing and managing the capabilities of sensor networks. In this paper, we introduce a system model that enables end-to-end secure connectivity between low-power IoT devices and UAVs, that helps to manage the data processing tasks of heterogeneous wireless sensor networks. The performance of proposed solution is analyzed by using simulation results. Moreover, in order to demonstrate the practical usability of the proposed solution, the prototype implementation is presented using commercial off-the-shelf devices.

\end{abstract}


\section{Introduction}

The world of Internet of Things~(IoT) allows the data retrieved by multiple smart devices to transmit to different cloud servers, in order to build a more comprehensive picture of the whole ecosystem~\cite{al2015internet}. However, this needs seamless connectivity between the IoT sensor and the central cloud (i.e., end-to-end connectivity), which is hard to be guaranteed every time~\cite{bello2016intelligent}. Edge computing paradigms (eg., Fog computing, cloudlets) have appeared to bridge the gap between remote clouds and the IoT devices~\cite{shi2016edge,porambage2018survey}. 

Typically, low power IoT devices are equipped with unlicensed band short-range radio access technologies, including Bluetooth Low Energy (BLE), HaLow, ZigBee, and Smart Utility Networks (SUNs)~\cite{xia2018radio}. BLE is a commonly used stateless protocol where the requests are independently transmitted. Moreover, BLE allows a flexible topology that can be adjusted to accommodate in many applications~\cite{porambage2019managing}. The IoT sensor nodes that use standalone BLE for communication purposes require dedicated nodes that serve as local gateways~(GWs) to provide back-end connectivity with the central cloud servers. In our previous work, we presented a mobile-based relay assistance solution for establishing secure End-to-End~(E2E) connectivity between low-power IoT sensors and cloud servers without using a dedicated gateway~\cite{porambage2019managing,manzoor2018mobile}.

Recently, Unmanned Aerial Vehicle~(UAV) systems or drones received great attention for autonomous monitoring applications~\cite{zeng2016wireless}. However, to the best of our knowledge there is still no sufficient work available to show the synergies between UAV systems and low power IoT devices and their cooperative applicability with practical implications in sensor networks. To describe the concept of integrating end-to-end secure connectivity between low power IoT devices and UAVs in a real world problem, we contemplate a smart agriculture use case which consists of deployed wireless sensors. Therefore, we propose this system to a remote agricultural site which is difficult to reach and monitored by humans frequently. A drone can be sent to the farm site to collect sensed data and perform actions (e.g. apply water or fertilizer, activate crop monitoring sensors). These actions can be activated based on the data received by the sensors, which are then processed already at the drone. The user can control the drone, send control commands, or program the flight route whenever needed. Similar setup can be used for environmental monitoring, disaster detection, anomaly detection in sensor networks or mobile crowd sensing applications. This setup can be extended for dew computing~\cite{ray2018introduction} based applications.

We present a novel edge computing based architecture which entails an UAV to assist data retrieving, data processing, and data management in a heterogeneous Wireless Sensor Network~(WSN) located in a remote area. The proposed system supports edge computing architecture to provide both insight and remote processing capabilities to the remotely located IoT sensor networks. Instead of using mobile phone as the relay in~\cite{porambage2019managing}, here we use a drone accompanied by a Raspberry Pi to perform as an edge processor. We have proven the viability of the proposed solution by providing the comprehensive analysis of the simulation results. Finally, a prototype implementation is presented using commercial off-the-shelf devices to demonstrate the practical viability.


The remainder of the paper is organized as follows: Section~\ref{sec:archi} describes the proposed  architecture and the communication protocols. Section~\ref{sec:num} and~\ref{sec:prototype} present the simulation and implementation results of the proposal. Finally, Section~\ref{sec:conclusions} concludes the paper by drawing the future research directions.

\section{The Proposed Architecture}
\label{sec:archi}
\subsection{System Model}
\begin{figure}[ht]
\centering
    \includegraphics[width=0.48\textwidth]{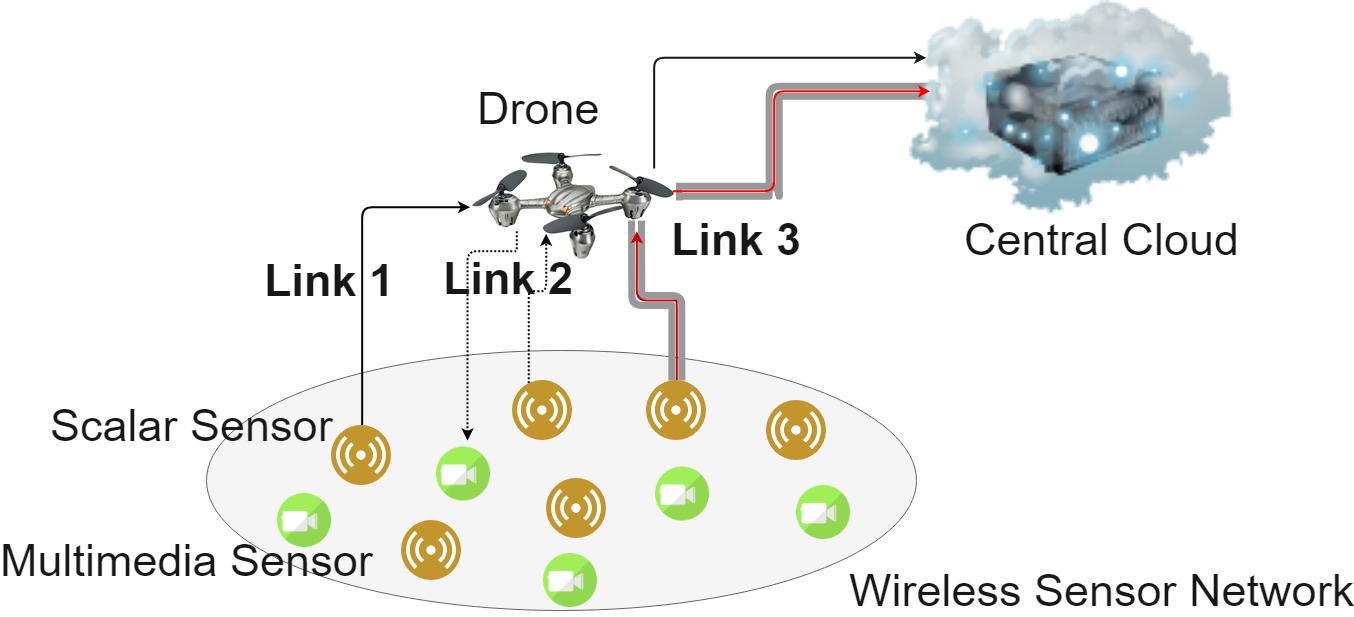}
  \caption{Use case scenario}
  \label{fig_systemmodel}
  \vspace{-3mm}
\end{figure}

As illustrated in Figure~\ref{fig_systemmodel}, we consider heterogeneous IoT devices which contain diversified sensor nodes with different capabilities whereas the drone is performing as the edge server. It will perform as a lightweight server and serves one sensor at a time. Drone can provide end-to-end connectivity from the sensor to cloud when there is Internet connectivity or allow offline accessibility otherwise. The drone will perform on site data processing and take the control functions of the IoT sensor network. These can be sensor nodes that monitors the environmental conditions and collect data or the actuator devices that perform given actions based on the received commands. We consider three types of links that can exist in this particular setting. 

\begin{itemize}
    \item \textbf{Link 1:} Sensors collect data and send to the central cloud via the drone, when the drone is returned to home location. Drone can access and process the retrieved data.
    \item \textbf{Link 2:} Sensors collect data and send to drone for further processing. Drone controls the actuator nodes based on the retrieved data. No data is transferred to the central cloud. As in edge computing paradigm, the drone will provide offline data accessibility to the actuator nodes, even though without real-time access to the central cloud.
    \item \textbf{Link 3:} Sensors collect data and send to the central cloud via the drone in the encrypted format. Drone is acting as a relay and cannot decrypt the data. Data offloading from the drone to cloud can be performed once the drone returns to home location.
\end{itemize}

\subsection{Communication Protocol}

Two communication protocols are used in the proposed architecture. First protocol is used to upload the data from sensor node to the drone and second protocol is used to download the data from the drone to the sensor node.

\subsubsection{Data upload process}

\begin{figure}[h]
\centering
    \includegraphics[width=0.48\textwidth,height=0.35\textheight]{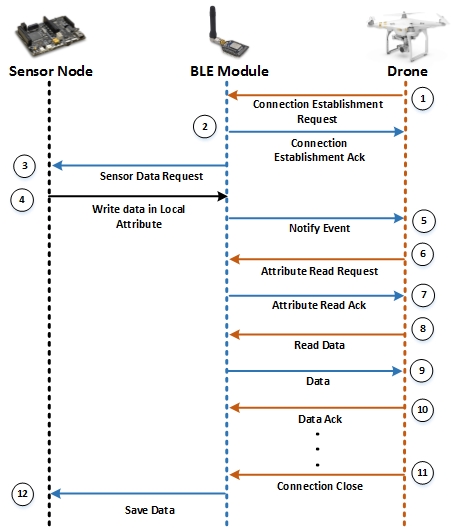}
  \caption{Communication Protocol for data upload process}
  \label{fig_protocol1}
  \vspace{-4mm}
\end{figure}

Figure~\ref{fig_protocol1} illustrates the message sequence between the sensor node, BLE module and the drone for the data upload process. Drone initiates the connection establishment process by sending connection request to the BLE module of the sensor node. Then BLE module acknowledges the connection request and connects with the drone. Next, BLE module requests data from sensor node, which has already been saved inside the sensor node. Then, sensor node writes data on BLE module’s local attribute. After that, BLE module notifies the drone that it has data to be transferred to the drone. Drone then sends attribute read request to the BLE module. Once the attribute request is acknowledged by BLE module, drone starts reading data and sends acknowledgment to the BLE module to ensure a reliable communication between the drone and the BLE module. This process continues until the drone retrieves required information and then the drone terminates the connection. Once the drone has terminated the connection, BLE module sends a command to the sensor node to save the data until it makes another connection with the drone in its next flying cycle.

\subsubsection{Data download process}

\begin{figure}[h]
\centering
    \includegraphics[width=0.38\textwidth,height=0.25\textheight]{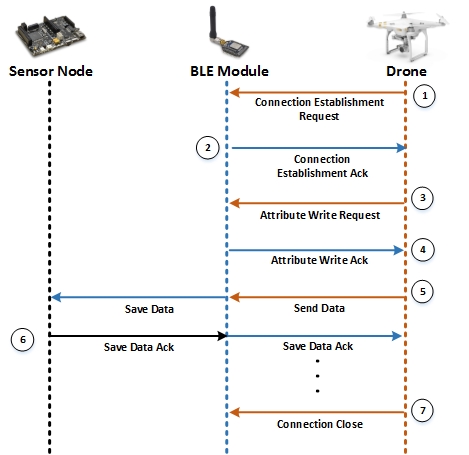}
  \caption{Communication Protocol for data download process}
  \label{fig_protocol2}
  \vspace{-5mm}
\end{figure}

Figure~\ref{fig_protocol2} illustrates the message sequence between drone, BLE module and the sensor node during data download. Similar to the upload procedure, drone initiates the connection process by sending connection request to the BLE module of the sensor node. BLE module acknowledges this request and connects with drone. Then the drone begins data download by requesting attribute writing permission from BLE module. Afterwards, BLE module acknowledges the request from the drone and starts retrieving data from drone. At the same time, BLE module sends the command to the sensor node to save those data in flash memory. Drone also receives an acknowledgement so that it knows the data has been successfully transferred.  After the successful completion of data transfer, drone terminates the connection.

\section{Numerical Analysis}
\label{sec:num}

To measure the performance of the proposed system, we implemented a simulation using MATLAB simulator. Simulation steps are explained below. Results of the simulation provide the number of sensor nodes, for which the drone can fly and collect its data, during one flying cycle. 

\begin{figure}[ht]
\centering
    \includegraphics[width=0.48\textwidth,height=5cm]{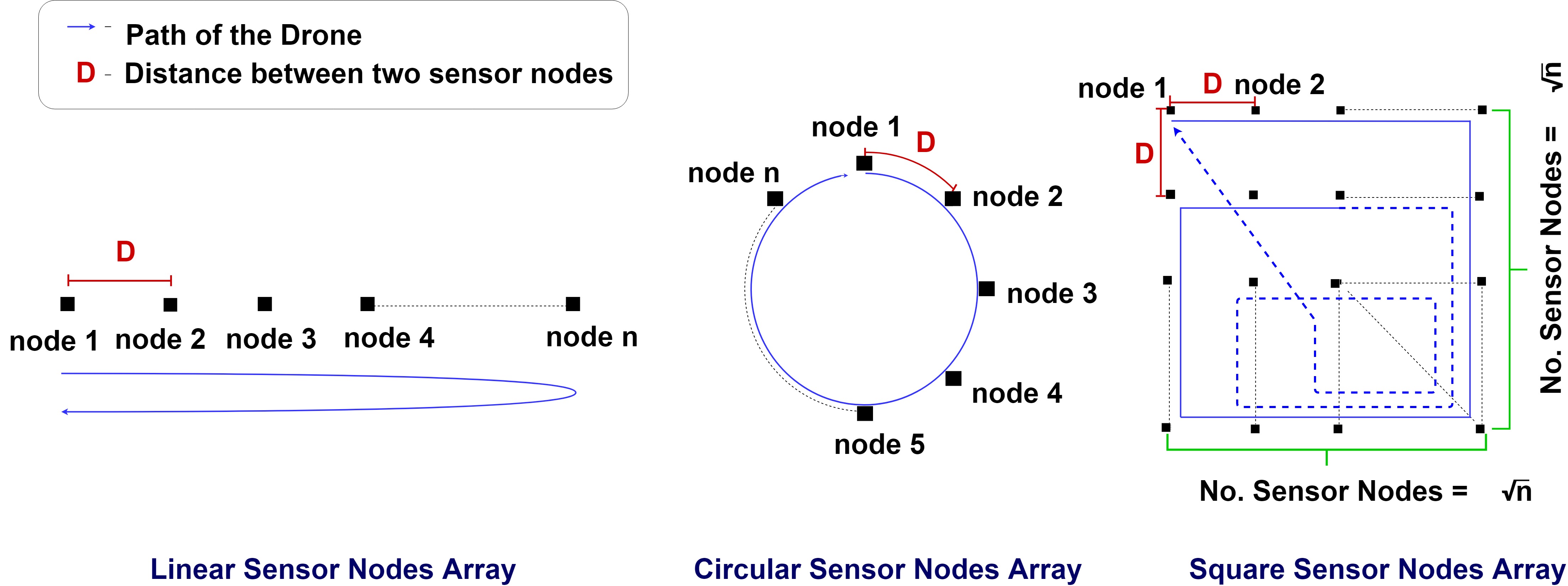}
    \caption{Orientation of Sensor Arrays}
  \label{fig_array}
   \vspace{-2mm}
\end{figure}

In the simulation setup, we considered three main sensor node arrays as depicted in Figure~\ref{fig_array}, which can be practically deployed in the environment. Therefore we contemplated the linear, circular and square orientations of sensor node arrays. For each array, we observe the number of sensor nodes in the array supported during drone’s one flying cycle, assuming that the drone starts flying from the first sensor node of the array. 

We consider that the sensor node array is comprised of $n$ sensor nodes as depicted in Figure~\ref{fig_array}. For the square sensor node array i. e \((\sqrt{n}\times \sqrt{n})\)  where we consider \((\sqrt{n})\) as an odd number.

Following (3), (4) and (5) equations illustrate the relationship between drone’s maximum hovering time and number of sensor nodes that can be covered during data collection procedure. Those equations were derived using the general flying time equation depicts in equation (1), where  \(n\) is the number of sensor nodes in a particular array.

\begin{equation}
 T_{F}   = (n\times T_{W}) + \frac{Distance\, between\, all\, nodes}{S}
\end{equation}

\begin{equation}
  T_{W} = \frac{T_{P}}{60}+\frac{T_{DL}}{60} + \frac{R_{S}(T_{C}+T_{F})}{R_{BLE}}
\end{equation}

\begin{equation}
   n_1 = \frac{T_{F} + \frac{D}{30S}}{T_{W} + \frac{D}{30S} }
\end{equation}

\begin{equation}
   n_2 = \frac{T_{F}}{T_{W} + \frac{D}{30S} }
\end{equation}

\begin{equation}
   n_3 = \Bigg\{\frac{\frac{-D}{\sqrt{2}}+\sqrt{\frac{D^2}{2}+ 4A\times \Big((1+\frac{1}{\sqrt{2}})D+60ST_{F}\Big)}}{2A}\Bigg\}^2
\end{equation}

\begin{equation}
   A = (60ST_{W})+D
\end{equation}

Where,
\(T\textsubscript{W}\) is the drone's waiting time near a sensor node. \(n\textsubscript{1}\), \(n\textsubscript{2}\), \(n\textsubscript{3}\) are the number of sensor nodes in linear sensor node array, circular sensor node array and square sensor node array respectively that can be covered during data collection procedure. \(T\textsubscript{F}\) is the total flying time of the drone, assuming that the drone starts flying from the first sensor node of the array. \(T\textsubscript{C}\) is the battery charging time of the drone, \(T\textsubscript{P}\) is the data processing time of Raspberry Pi, \(T\textsubscript{DL}\) is the data download time from Raspberry Pi to sensor node and \(S\) is the speed of the drone. \(R\textsubscript{BLE}\) is the data transfer rate between Raspberry Pi and sensor node, \(R\textsubscript{S}\) is the data generation rate of sensor node and \(D\) is the distance between two sensor nodes. \(A\) is used for simplify the equation (5).

Table~\ref{table:gen_par} summarizes parameters we used in the experiments in order to obtain the simulation results. Moreover, we have evaluated how significantly each parameter contributes to decide the number of sensor nodes serviced by the drone. 

\begin{table}[ht]
  \begin{center}
    \caption{ General simulation parameters}
    \label{table:gen_par}
    \begin{tabular}{|p{5.5cm}|p{2.5cm}|} 
     \hline

      \textbf{Parameter} & \textbf{Value}\\

      \hline
    Total flying time \((T\textsubscript{F})\) & 25 minutes~\cite{inet}\\
    Battery charging time of the drone \((T\textsubscript{C})\)  & 90 minutes~\cite{inet} \\
        Data processing time of the drone \((T\textsubscript{P})\) & 1 second \\
         Data download time from drone to node \((T\textsubscript{DL})\) & 10 seconds \\
        Data generation rate of the sensor node \((R\textsubscript{S})\) & 20 B/min~\cite{inetwasp} \\
         BLE data rate between drone and node \((R\textsubscript{BLE})\) & $10.5\times 10^4$B/min~\cite{inetBLE} \\
         Speed of the drone \((S)\) & 12 m/s \\
         Distance between two sensor nodes  \((D)\) & 100 m \\
      \hline
    \end{tabular}
  \end{center}
\end{table}
 
\begin{figure*}
\centering
    \includegraphics[width=16cm,height=18cm]{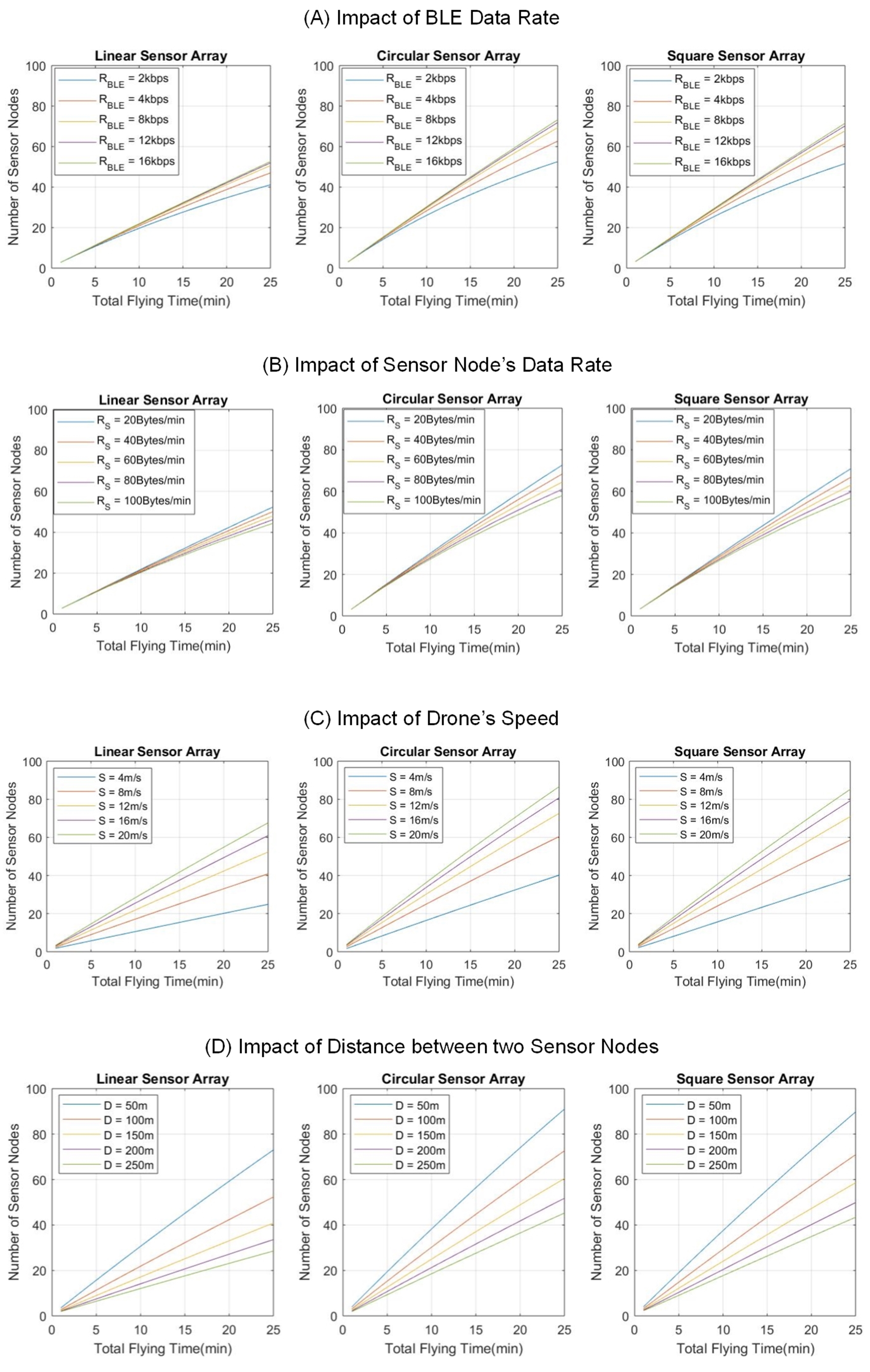}
  \caption{Simulation Results}
  \label{fig_plots}
  \vspace{-2mm}
\end{figure*}

 Figure~\ref{fig_plots} illustrates the results of the experiment obtained using MATLAB simulator. 
\subsection{Impact of BLE Data Rate}

In our proposed system, we use BLE communication protocol between the drone and the sensor nodes. Data rate of BLE communication has a significant impact on waiting time of the drone near the location of sensor node. Waiting time of the drone at a sensor node decreases as BLE data rate increases. Therefore, with lower waiting time spent at a particular sensor node, drone can collect data from more sensor nodes throughout its maximum flying time.

In order to illustrate the relationship between the BLE data rate and the number of sensor nodes supported at each BLE data rate, we considered different values for BLE data rate and executed the simulation. We considered BLE data rate\((R\textsubscript{BLE})\) as 2 kbps, 4 kbps, 8 kbps, 12 kbps, 16 kbps while keeping the other parameters fixed. These parameters are depicted in Table~\ref{table:gen_par}. This experiment has been carried out for all three sensor node arrays i.e. linear, circular and square. 

Figure~\ref{fig_plots}(A) illustrates the results of the simulation. It is clear that, with a higher BLE data rate, drone is capable to collecting data from more nodes and this observation is valid for all three sensor node arrays. Results further demonstrated that circular and square arrays yield better performance than the linear array of sensor nodes.

\subsection{Impact of Sensor Data Rate}

Sensor node data rate, also has a substantial impact on the performance of the system. With lower sensor node data rate, drone can collect data from more sensor nodes than having higher data rates. This behaviour is opposite to the behaviour we considered for BLE data rate.  When the sensor data rate is low, the amount of data that the drone has to collect is low, thereby requiring a less amount of time to retrieve the data. Figure~\ref{fig_plots}(B) illustrates the simulation results for this scenario. Here we have changed sensor node's data rates\((R\textsubscript{S})\) as 20 Bytes/min, 40 Bytes/min, 60 Bytes/min, 80 Bytes/min, 100 Bytes/min while keeping other parameters in Table~\ref{table:gen_par} fixed. Also we assumed one in-built sensor generates 20 Bytes/min. In this case also, circular and square arrays provided better performance than linear sensor node array.

\subsection{Impact of Drone's Speed}

In our prototype implementation, we used DJI Phantom 3  SE drone which is capable of flying at maximum 16 m/s ~\cite{inet}. For the simulation experiments we used 12 m/s as average speed\((S)\) considering the time taken to accelerate and decelerate. By using a drone with a higher maximum flying speed, the performance of the system can be enhanced. In order to validate this phenomenon, we carried out our simulation experiments from lower drone speeds to higher speeds. We changed speed values as 4, 8, 12, 16, 20 m/s  while keeping other parameters fixed as in Table~\ref{table:gen_par}. Figure~\ref{fig_plots}(C) illustrates the simulation results for this scenario. According to Figure~\ref{fig_plots}(C), drones with higher speed are capable of collecting data from more sensor nodes than drones with lower speeds. In this case also, circular and square array provided better performance  compared to linear array.

\subsection{Impact of Distance Between Sensors}

In order to analyze how the distance between two sensor nodes affects the number of sensor nodes supported, we used 50 m, 100 m, 150 m, 200 m, 250 m as distance\((D)\) and carried out the simulations. In this case also we kept other parameters fixed as in Table~\ref{table:gen_par}. Based on the results as depicted in Figure~\ref{fig_plots}(D),  the lower the distance between two sensor nodes, the higher the number of sensor nodes supported by the drone  during data collection. Moreover, circular and square arrays provided better system performance than the linear array.





\section{Prototype Implementation}
\label{sec:prototype}

To establish the prototype implementation we used five Waspmotes with BLE modules, a Raspberry Pi 3 and a DJI phantom 3 SE ~\cite{inet} drone as core instruments of the setup.

\begin{figure}[ht]
\centering
    \includegraphics[width=0.5\textwidth]{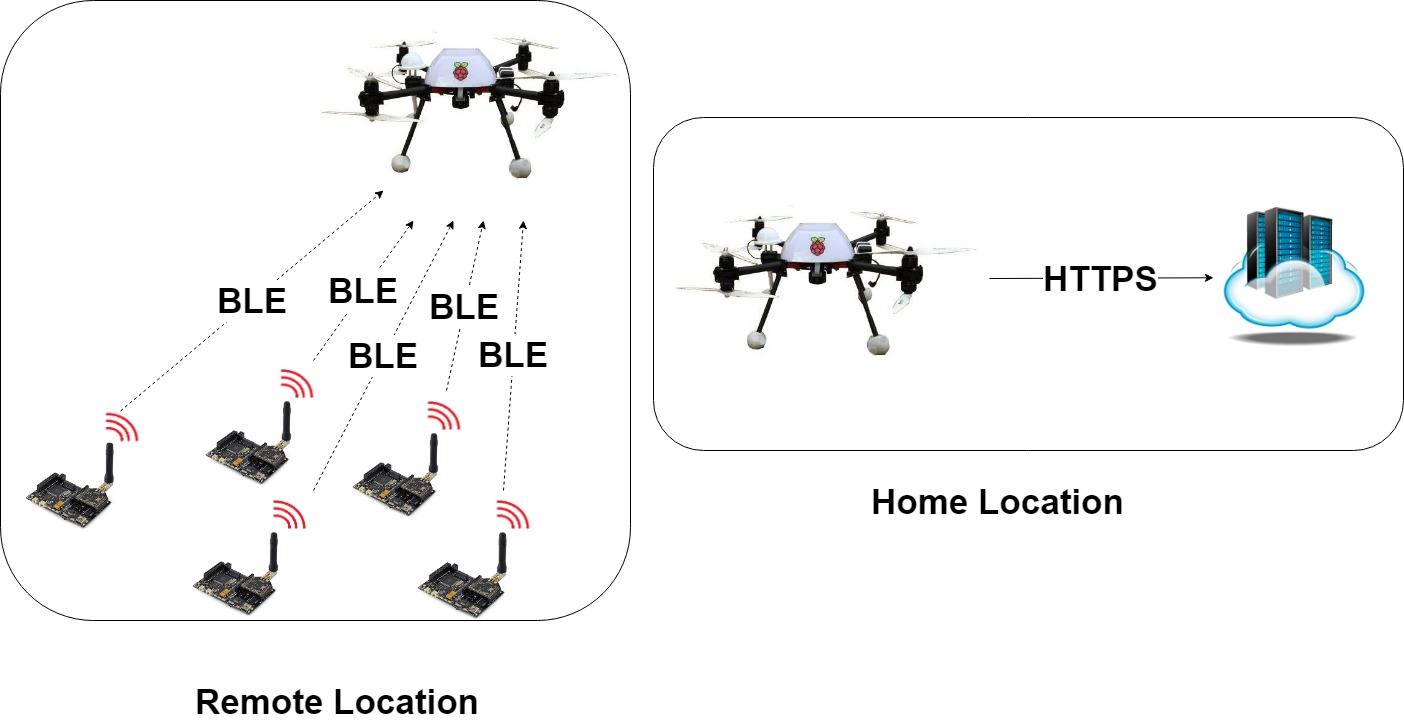}
  \caption{Prototype}
  \label{fig_prototype}
\end{figure}

Figure~\ref{fig_prototype} illustrates system model of the prototype implementation. For the prototype implementation, we first implemented the Waspmote sensor network. To setup the BLE communication capability of Waspmote we used Waspmote BLE module introduced by LIbelium ~\cite{inetwasp}. For the implementation, we considered Wapmote's in-built accelerometer generated data. This data is then transferred to Raspberry Pi over BLE. Raspberry Pi is coupled with the drone. Generally, Raspberry Pi 3 is powered by a +5.1V micro USB supply and recommended input current is 2.5A~\cite{inetrasp}. To cater these specifications we used a portable external power source coupled with Raspberry Pi. This prototype implementation was basically based on two scenarios, real time and non-real time.

To initialize the connection between Raspberry Pi and Waspmote sensors, we defined Waspmote as the master and Raspberry Pi as the slave. Therefore Waspmote sends advertising packets to Raspberry Pi through advertising attribute and then Raspberry Pi sends the connection request to Waspmote to establish the connection. Moreover, Raspberry Pi uses the MAC addresses of Waspmotes to establish the connection.

The real time scenario has implemented to collect the data when the drone is within the range of sensor network. Hence, in the real time scenario we created a local attribute in Waspmote for data transferring process. Therefore we programmed a handler in BLE user services profile of Waspmote to send the accelerometer sensed data to Raspberry Pi attached to the drone.

Non-real time scenario has proposed and implemented to maintain the continuity of data collection while drone was away from the sensor network.  In the non-real time scenario Waspmote generates the accelerometer data and saves them into a text file in its SD memory. When,  the drone arrives Waspmote sends those text files through pre-defined handler as in the real time scenario. 

\begin{figure}[ht]
\centering
    \includegraphics[width=0.5\textwidth]{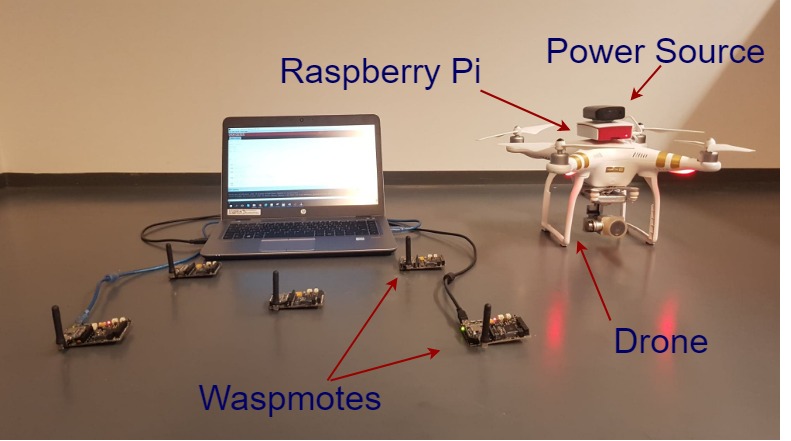}
  \caption{Prototype Implementation}
  \label{fig_photo}
  \vspace{-3mm}
\end{figure}

After extracting the sensed data (real time and non-real time) from Waspmotes, Raspberry Pi saves all sensed data into a SQLite database. Hence when the drone comes to its home location, Raspberry Pi sends those saved data from SQLite database to the cloud server over HTTPS protocol. The prototype setup with equipment is illustrated in Figure~\ref{fig_photo}.

\subsection{Cloud Integration}

The last part of the implementation was to deploy the cloud storage server. As the drone arrives at the docking station after extracting the data from sensors, Raspberry Pi 3 coupled with the drone connects to 802.11 WLAN router with internet access; to upload the locally saved sensor data.  The cloud server is implemented on Google Firebase\footnote{https://firebase.google.com/} where Raspberry Pi first authenticates itself using its private key. After authentication, Raspberry Pi starts uploading the sensed data to cloud and stores in the database. The Firebase Real-time Database is a cloud-hosted database. Data is stored as JSON and synchronized in real-time to every connected client. Moreover, Firebase uses HTTPS connection over Transport Layer Security for secure communications between Raspberry Pi and cloud server along with real-time database security. The database provides a flexible, expression-based rules language, to define how your data should be structured and when data can be read from or written to the database. It is capable of defining who has access to what data, and how they can access it. After successful transferring of the data, Raspberry Pi deletes the stored data from the local database.

\subsection{Performance Results}

We used handler 0x0038 from BLE user services profile of Waspmote to send the accelerometer sensor data to the Raspberry Pi. Handler 0x0038 is capable of creating a reliable BLE communication between Raspberry Pi and Waspmote, because it caters the master with read/write and indication operations. Maximum data size that can be transferred during a particular transmission attempt is 20 bytes. Waspmote sends those accelerometer sensor data in an encoded format which has to be decoded at Raspberry Pi to make it readable. After the decoding process, it extracts to 112 Bytes at Raspberry Pi. Hence, we consider this 112 Bytes as maximum data that can be send within one particular transmission. To obtain the performance results, we fixed the drone's flying altitude as 10 m above the ground level. 

For the data download scenario (i. e. the transmission from Raspberry Pi to Waspmote) we used handler 0x0029. Here we considered 10 seconds time duration to send the data packets over BLE from Raspberry Pi to Waspmote. The following Table~\ref{table:exp_par} summarizes the results obtained from the experiment.

According to the results depicted in Table~\ref{table:exp_par}, Raspberry Pi to Waspmote data download rate is higher than Waspmote to Raspberry Pi data upload rate. With the buffer size limitation in Waspmote, Raspberry Pi is only allowed to send a maximum of 74 data packets (each 20 Bytes) to Waspmote. This maximum data transmission causes to fill the buffer size at Waspmote and then stops retrieving data from Raspberry Pi.

\begin{table}[ht]
  \begin{center}
    \caption{ Experimental Data Rates}
    \label{table:exp_par}
    \begin{tabular}{|c|c|} 
     \hline

      \textbf{Data Rate Parameter} & \textbf{Value}\\

      \hline
     
 Data transfer rate from Waspmote to Raspberry Pi \(R\textsubscript{UL}\)  & 672 Bytes/min \\
Data transfer rate from Raspberry Pi to Waspmote  \(R\textsubscript{DL}\) & 148 Bytes/s \\
      \hline
    \end{tabular}
  \end{center}
\end{table}

\section{Conclusions}
\label{sec:conclusions}

The widespread of low-power IoT technologies and their sensing and monitoring applications may demand processing power at the edge of the network, rather than accessing the remote central cloud. Throughout this paper, we have addressed how to provide edge processing power in a similar manner as in edge computing architecture, for the low-power BLE sensors with the help of UAVs such as drones. We have described the protocol in detail along with the simulation and prototyping results. The key findings of the simulations are as follows: the number of served sensor nodes by the drone is directly proportional to the speed of the drone; The distance between two sensor nodes is the most dominant parameter that defines the overall performance of the system; Circular and square array type sensor node topologies always outperform the linear array topology. The prototyping caused some problems due to the hardware limitations of the Waspmotes.

We intend to extend the research by implementing a fully working prototype and measuring energy consumption in an actual WSN where the sensors are deployed in a higher density. The current experimental results are taken in an ideal situation assuming a constant drone speed by taking the wind speed as zero. In the future, we plan to investigate how the drone speed and the wind speed will impact on the proposed system model. 

\section*{Acknowledgement}
This work is supported by Business Finland in Towards Digital Paradise, Academy of Finland in SECUREConnect, 6Genesis Flagship (grant no. 318927), 5GEAR projects, and European Union in RESPONSE 5G (Grant No: 789658) project. A. Gurtov was supported by CENIIT project 17.01. The authors would also like to acknowledge the contribution of the COST Action CA15127 (RECODIS) and CA16226 (SHELD-ON) projects.


\bibliographystyle{IEEEtran}
\bibliography{references}

\vfill
\vspace{0.4cm}

\end{document}